\documentclass[a4paper,11pt]{article}
\usepackage{pos}
\usepackage{xspace}
\usepackage[capitalize]{cleveref}
\graphicspath{{fig/}}

\title{Attenuation Models for Extensive Air Showers Derived from Simulations}
\ShortTitle{Attenuation Models for Extensive Air Showers}

\author{Fiona Ellwanger}
\author*{Darko Veberi\v{c}}

\affiliation{Institute for Astroparticle Physics, Karlsruhe Institute of Technology \includegraphics[height=1.55ex]{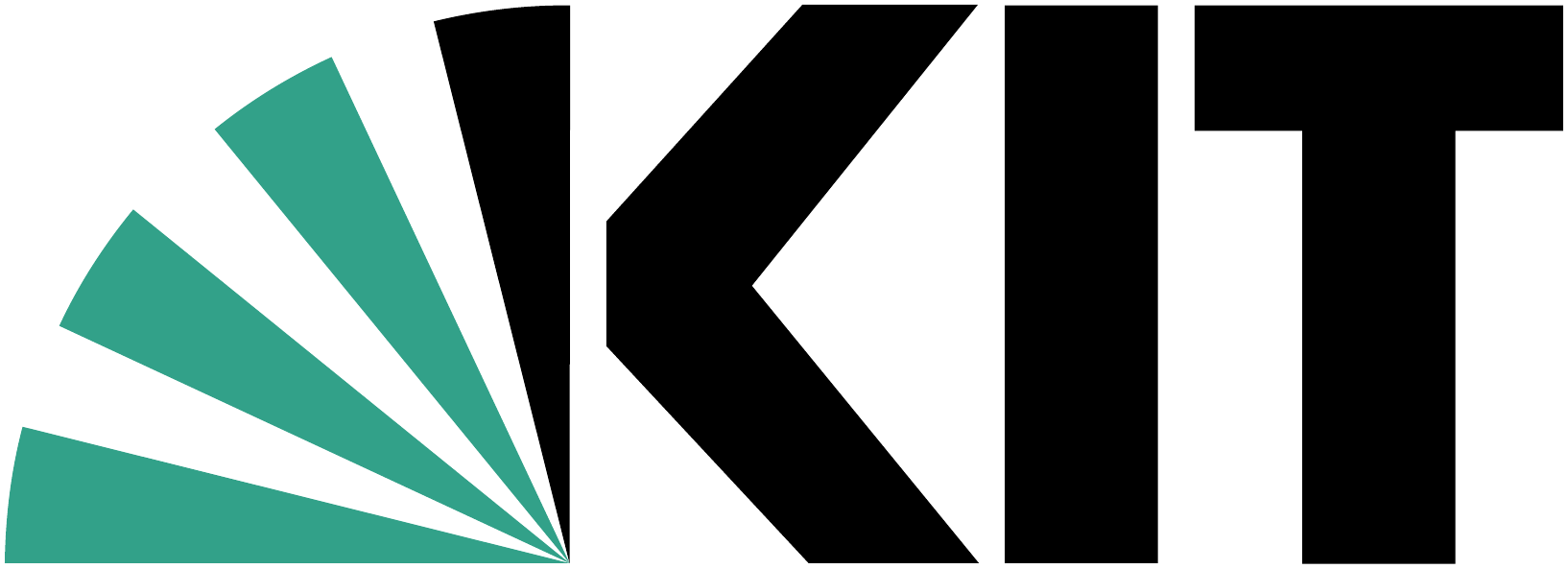}, Karlsruhe, Germany}

\emailAdd{fiona.ellwanger@kit.edu}

\abstract{
At ultra-high energies, the flux of cosmic rays is too low for direct measurements to be meaningful.
When a cosmic ray enters the atmosphere, it initiates an extensive air shower, producing a cascade of secondary particles that propagate toward the ground.
Large arrays of surface detectors are used to measure these secondary particles upon arrival.

The signal detected at a specific reference distance from the shower core serves as a proxy for the shower size and, consequently, as a reliable estimator of the energy of primary cosmic ray.
However, shower development is influenced by attenuation effects: measured signals at the ground depend on the amount of traversed atmospheric density (column density) through which the shower evolves.
Since the column density varies with the inclination of the shower, it is important to account for these attenuation effects to ensure accurate energy estimation.

In this study, we derive physics-and-geometry-based functional forms to describe attenuation and propose appropriate expansion terms using simple one-dimensional shower-development models, incorporating one or two main particle-cascade components.
We then evaluate the applicability and effectiveness of these functional forms using a Monte-Carlo dataset that includes various primary cosmic-ray particles.
By directly calibrating the the shower size derived from ground signals to the Monte-Carlo energy, we characterize attenuation behavior across different primary particles, assess the energy dependence of attenuation, and quantify systematic uncertainties introduced by different functional forms.
}

\ConferenceLogo{PoS_ICRC2025_logo}

\FullConference{%
39th International Cosmic Ray Conference (ICRC2025)\\
15 -- 24 July 2025\\
Geneva, Switzerland}

\def\Offline{\mbox{$\overline{\textrm{Off}}$\hspace{.05em}\protect\raisebox{.4ex}{$\protect\underline{\textrm{line}}$}}\xspace}
\def\eq#1{\begin{align}#1\end{align}}

\newcommand{\fatt}{f_\text{att}}
\newcommand{\rr}[1]{\llbracket{#1}\rrbracket}

\begin{document}
\maketitle

%
%
%

\section{Motivation}

Even though we use in our Monte-Carlo study an example of the surface detector of the Pierre Auger Observatory with a spacing of 1500\,m (SD-1500), in general methods presented below apply equally well to all other measurements of extensive air showers using surface detector arrays.

A reconstruction of an event typically combines individual signals in the detector stations with an intention to derive a suitable proxy for the \emph{shower size at the ground}, e.g.\ by fitting a lateral distribution function to the observed signals.
The shower size at the ground, $S_\text{gr}$, can then be approximated, for example by the expected signal $S_\text{gr}(r_\text{opt})$ at an optimal distance $r_\text{opt}$~\cite{newton} from the shower axis.
For the SD-1500 of the Pierre Auger Observatory the optimal distance is $r_\text{opt}=1000$\,m.

Depending on the shower geometry, the ``true'' or ``intrinsic'' shower size or simply just \emph{shower size} $S_\text{sh}$ gets attenuated so that the measured signal at the ground $S_\text{gr}$ depends on the zenith angle $\theta$ of the incoming direction.
%
%
For example, to describe the attenuation, the Pierre Auger Collaboration is using a cubic polynomial function, see~\cref{e:cubic}.
Using a very generic and flexible ansatz, like the polynomial, might be enough for quick elimination of attenuation effects.
However, using a polynomial expansion around some chosen reference point implicitly assumes (a) that the attenuation function is only slowly changing in its vicinity and (b) that its shape can be well described by the lowest orders of power for arguments far from the reference.
This might not be the case, especially considering how fast the atmospheric column density can increase with the inclination of the shower.
For these reasons, in the following sections we will try to derive more practical functional forms of attenuation based on some simple physical assumptions about the shower development, which, hopefully, will have better extrapolation properties when used outside of the specific fit range.

\section{Simulations}

\begin{figure}
\centering
\includegraphics[width=\linewidth]{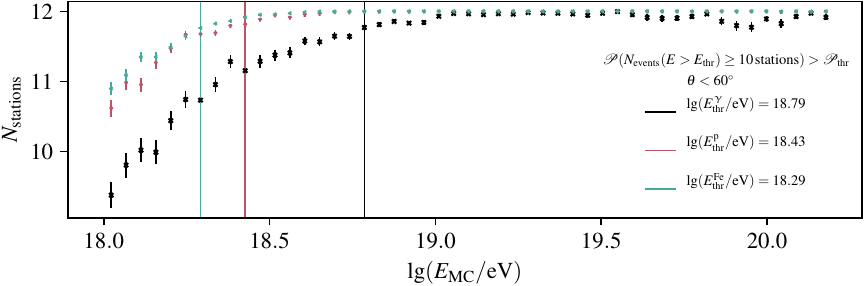}
\caption{Number of triggered stations in the ring at 1000\,m around the shower axis for simulated iron (green), proton (red), and photon-induced showers (black).
The vertical lines indicate the corresponding threshold energies defined as the energy at which a fraction of $\mathcal{P}_\text{thr}$ of events have at least 10 triggered stations in the ring of 12 simulated stations.
For proton and iron-induced showers we chose $\mathcal{P}_\text{thr}=99.9\%$.
Since the efficiency for photon-induced showers is not monotonically increasing with the energy, we lower the threshold in the case of photons to reach only $\mathcal{P}_\text{thr}=99\%$.}
\label{f:efficiency}
\end{figure}

We will compare the observed attenuation of simulated proton, iron, and photon-induced extensive air showers.
The showers were simulated using \textsc{Corsika}~\cite{corsika} version 7.7420 and the \textsc{EPOS-LHC} hadronic interaction model~\cite{epos} and were simulated in an energy range from 1 to 160\,EeV with a spectral index of $-1$ and with zenith angles $\theta$ between 0 and $60^\circ$ following a uniform distribution in $\sin^2\theta$.
In total, approximately 20\,000 showers have been used for each primary particle.
For the detector simulation we use the setup of water-Cherenkov detectors for the Pierre Auger Observatory using the \smash[t]{\Offline} software package~\cite{offline}.
In analogy to the energy reconstruction used in the Pierre Auger Collaboration, we exploit the signal measured at 1000\,m from the shower axis as a suitable proxy for the shower size at the ground.
In the simulation, 12 virtual water-Cherenkov stations are placed in a ring around the shower axis at a distance of exactly 1000\,m.
The shower size $S_\text{gr}$ at the ground is then estimated by taking signals $S_{1000}^i$ in these stations $i$ and fitting them to the lowest possible order in azimuthal asymmetry as
\eq{
  S_{1000}^i = S_\text{gr} \, (1 + \alpha\,\cos\phi_i^\text{sp}),
\label{e:dense_fit}
}
where $\phi_i^\text{sp}$ is the azimuth angle of the station $i$ in the plane perpendicular to the shower (\emph{shower plane}) and where $\alpha$ is the amplitude of the asymmetry.
In this way we take into account the asymmetry of the shower at the ground and still obtain a good estimate of $S_\text{gr}$ even in the case of several non-triggered stations.
Nevertheless, when all 12 stations are present, $S_\text{gr}$ resulting from \cref{e:dense_fit} is equivalent to a mean over all $S_{1000}^i$ values.
For the fit we require presence of at least 10 triggered stations and restrict the fit only to the energies above which $\mathcal{P}_\text{thr}$ fraction of events satisfy this requirement, see \cref{f:efficiency}.
For proton and iron-induced showers we chose $\mathcal{P}_\text{thr}=99.9\%$. 
Since for photon-induced showers the efficiency is not monotonically increasing with the energy, for photons we lower this requirement to $\mathcal{P}_\text{thr}=99\%$.

\section{Physically motivated forms of the attenuation function}

\paragraph{Atmosphere.}

For attenuation of showers -- as observed at the ground -- the decisive quantity is the amount of traversed matter i.e.\ slant column density.
In the static and isothermal approximation to the conditions of the atmosphere, the density of air $\rho$ depends only on height, $\rho(h)=\rho_\text{gr}\exp(-h/h_\text{s})$, where $\rho_\text{gr}$ is the air density at the ground level ($h=0$) and $h_\text{s}$ is the (assumed-constant) scale height of the atmosphere.
%
Assuming a flat Earth and parallel air density layers, we get $X_\text{gr}(\theta)=X_\text{gr}^\text{v}\sec\theta$, where $X_\text{gr}^\text{v}$ is the vertical column density.
The exact column density in a curved atmosphere is described by the Chapman function~\cite{vasylyev} but the deviation from the flat approximation only becomes important for very inclined showers, which is beyond the required accuracy for our models in this work.

\subsection{Model with two components.}

Since the response of either water-Cherenkov or scintillator detectors to secondary shower particles is dominated by the electromagnetic and muonic component, we will from now on assume that showers consist only of these two components.
The observed attenuation behavior of the whole shower is then a superposition of the individual attenuation of the two components.


\paragraph{Electromagnetic component.}

Let us assume that the measured electromagnetic (em) signal at a certain point is proportional to the number of particles $N_\text{em} \propto \rho_\text{em}$ detected.
This number in turn depends on a specific stage of the longitudinal development of the shower, which again depends on the traversed column density $X$ until that point.
The dependence of the longitudinal particle density $\rho_\text{em}(X)$ in the electromagnetic cascade on the amount of traversed matter $X$ can be modeled with a Gaisser-Hillas function
\eq{
  \rho_\text{em}(\theta) \propto
  f_\text{GH} =
  \left(
    \frac{X-X_0}{X_\text{max}-X_0}
  \right)^{\frac{X_\text{max}-X_0}{\lambda}}
  \exp\left(
    -\frac{X-X_\text{max}}{\lambda}
  \right),
\label{e:gaisser-hillas}
}
where $X_\text{max}$, $X_0$, and $\lambda$ are the essential Gaisser-Hillas shower parameters~\cite{gh}.
We will use this form as a simplified approximation proportional to the signal at the depth $X$.
In doing so, we neglect, in this first stage of derivation, the muonic component of the showers.
We can thus write for the signal $S$ at a certain depth $X$,
\eq{
  S(X) = S_\text{max}\,f_\text{GH}(X; X_\text{max}, X_0, \lambda).
}
In this simple model, the particles in the shower continue producing secondary particles until they reach the critical energy $E_\text{crit}$.
This defines the point of maximum shower development, i.e.\ when $X=X_\text{max}$, and the number of particles in the shower reaches its peak at $N_\text{max}$.
A detector, approximated as a particle counter, would measure the highest signal, $S_\text{max}$, when positioned at $X=X_\text{max}$.
Since the energy of the primary particle is distributed among $N_\text{max}$ particles of energy $E_\text{crit}$ we can claim that the primary energy $E$ is
\eq{
  E \propto N_\text{max} \propto S_\text{max},
\qquad\text{and}\qquad
  E \propto \frac{S(X)}{f_\text{GH}(X; X_\text{max}, X_0, \lambda)}.
  \label{eq:energy-prop1}
}
We can now introduce a suitable reference depth $X_\text{ref}$ where $f_\text{GH}^\text{ref} = f_\text{GH}(X_\text{ref}; X_\text{max}, X_0, \lambda)$.
The primary energy is thus
\eq{
  E \propto
  \frac{1}{f_\text{GH}(X_\text{ref})} \,\frac{S_\text{gr}}{f_\text{GH}(X) / f_\text{GH}(X_\text{ref})},
  \label{eq:energy-prop2}
}
where the first term is just a constant and the zenith-angle dependence of the signal $S_\text{gr}$ at the ground or the ``attenuation'' $\fatt^\text{em}$ is thus described by
\eq{
  \fatt^\text{em}(X) =
  \frac{f_\text{GH}(X)}{f_\text{GH}(X_\text{ref})} =
  \left(
    \frac{X-X_0}{X_\text{ref}-X_0}
  \right)^{\frac{X_\text{max}-X_0}{\lambda}}
  \exp\left(
    - \frac{X-X_\text{ref}}{\lambda}
  \right).
\label{gha}
}
Inspired by this result let us first assume that $X_0$ is close to zero and can thus be safely neglected,
\eq{
\fatt^\text{em}(X) \approx
  \left(
    \frac{X}{X_\text{ref}}
  \right)^{X_\text{max}/\lambda}
  \exp\left(
    - \frac{X-X_\text{ref}}{\lambda}
  \right) =
  \left(
    \frac{\sec\theta}{\sec\theta_\text{ref}}
  \right)^\mu
  \exp(-\gamma(\sec\theta-\sec\theta_\text{ref}))
\label{exp2}
}
with $X=X_\text{gr}\sec\theta$, $X_\text{ref}=X_\text{gr}\sec\theta_\text{ref}$, and $\gamma=X_\text{gr}/\lambda$, where $\mu=X_\text{max}/\lambda$.
We can thus propose a new form for the attenuation function,
\eq{
  \fatt^\text{em}(\theta;\gamma,\mu) =
  \left(\frac{\sec\theta}{\sec\theta_\text{ref}}\right)^\mu
  \exp[-\gamma(\sec\theta-\sec\theta_\text{ref})] =
  \exp[ -\gamma\,x + \mu\,y],
}
with $x = \rr{\sec\theta}$ and $y = \rr{\ln\sec\theta}$ using the notation $\rr{z} = z - z_\text{ref}$.
Since $\gamma=X_\text{gr}/\lambda$ and $\mu=X_\text{max}/\lambda$ it is advantageous to use $\mu=\gamma-\delta$ as the fitting parameters in order to reduce the regression correlations.
This form thus changes into
\eq{
\fatt^\text{em}(\theta;\gamma,\delta) =
  \exp[-\gamma\,x +
       (\gamma-\delta)\,y].
  \label{fatt:EMAttenuation}
}

%
%

\paragraph{Muonic component.}

%
%
%
%
From the point of production muons are practically unimpaired by the atmosphere and decay with a typical decay length $\lambda_\upmu$.
%
%
%
%
%
From the muon profile in Fig.~1 of Ref.~\cite{atmo} where the total number of muons in a shower is given, we can convert the $X$ dependence into distance, obtaining $\lambda_\upmu\approx36$\,km.
We therefore have $\lambda_\upmu \gg h_\text{s}$ and we may alternatively treat muons as non-attenuating component, i.e.\ $\fatt^\upmu = 1$.

\paragraph{Total signal.}

To correctly model the attenuation behavior for the superposition of the two components we have to take into account the differences in the response of the detector to these components.
However, for simplification, we describe the total attenuation as a superposition of the attenuation of the two components
\eq{
  \fatt(\theta) =
    q_\text{em} \, \fatt^\text{em}(\theta) +
    (1 - q_\text{em}) \, \fatt^\upmu(\theta)
}
with the effective fraction of the electromagnetic component $q_\text{em}$.

\section{Direct Energy Calibration}

We can rewrite \cref{eq:energy-prop2} as $E = A'\,S_\text{gr}/f_\text{att}$, where $A'$ takes care of the proportionality, calibration, and units.
We also assume that the signal sizes are still dominated by the electromagnetic component, so that the proportionality in \cref{eq:energy-prop1} is not broken, which is also why the analysis is limited to zenith angles below $60^\circ$. 
The reference angle $\theta_\text{ref}$ is usually chosen close to the median of the distribution of zenith angles, resulting in a reference angle of around $38^\circ$.
%

As the uncertainties of the shower size are usually log-normally distributed we actually fit the relation between the logarithm of the shower size and primary particle energy.
For the signal-to-energy relation we use the model
\eq{
  \lg E = A + B\,\lg(S_\text{gr}/f_\text{att}),
\label{e:leny}
}
in which we allow also for an exponent $B\not=1$.
We take the form of the likelihood $\mathcal{L}$ for the fit from Ref.~\cite{likelihood},
\eq{
\mathcal{L} =
  \prod_i^{E_i>E_\text{cut}}
    \sum_k^\text{all events}
      \frac{1}{\sigma_{\lg E,k}\,\sigma_{\lg S,k}\,\epsilon_k} \,
      \exp\left(
        -\frac{\lg^2(E_i/E(S_k))}{2\sigma_{\lg E,k}^2}
        -\frac{\lg^2(S_i/S_k)}{2\sigma_{\lg S,k}^2}
      \right).
}
As we are going to use the ``true'' Monte Carlo energy for the calibration we have to take the limit $\sigma_{\lg E,k}\rightarrow0$.
In this limit the first Gaussian term becomes a $\delta$-function with which we can integrate out the inner sum and derive essentially a least-squares fit for the inverse of \cref{e:leny},
%
%
%
\eq{
\mathcal{L} =
  \prod_i^{E_i>E_\text{cut}}
    \frac{1}{\sigma_{\lg S,i}\,\epsilon_i} \,
    \exp\left(
      -\frac{\lg^2(S_i/S(E_i))}{2\sigma_{\lg S,i}^2}
    \right).
}
If we restrict the energy cut $E_\text{cut}$ to energies where the respective detector is fully efficient ($\epsilon_i = 1$) the only reason this is not simply a least-squares fit are the intrinsic shower fluctuations $\sigma_\text{in}$ which appear as an \emph{additional fit parameter}.
This intrinsic fluctuation accounts for the natural shower-to-shower variability that limits the precision of energy reconstruction beyond measurement uncertainties, $\sigma_{\lg S,i}^2=\sigma_{\lg S_\text{gr},i}^2 + \sigma_\text{in}^2$, where $\sigma_{\lg S_\text{gr},i}$ is derived from the fit to the ring of simulated stations.
%
We observe that ignoring the intrinsic fluctuations, i.e. fixing $\sigma_\text{in} = 0$, will lead to biased results.


The minimization is done without bounds on the fit parameters.
To ensure the conditions $\sigma_\text{in}>0$ and $q_\text{em}\in [0,1]$ they are wrapped in a softplus and expit\footnote{$\operatorname{softplus}(x)=\ln(1+e^x)$, $\operatorname{expit}(x)=(1+e^{-x})^{-1}$} functions, respectively.

\section{Comparison of fits}

Now we want to compare how well the following models describe the attenuation in simulations.
We start with a cubic function,
\eq{
    f_{abc} = 1 + x(a + x(b + x c)),
\quad\text{where}\quad
    x = \rr{\sin^2\theta},
    \label{e:cubic}
}
and compare the results with our simplified two component model,
\eq{
    f_{\delta,\gamma,q_\text{em}} = q_\text{em}\,\exp(-\gamma\,x + (\gamma - \delta)\,y) + (1-q_\text{em}) \, 1 
\quad\text{where}\quad
    x = \rr{\sec\theta}, \,
    y = \rr{\ln\sec\theta},
}
motivated by which we will suggest an exponential cubic function,
\eq{
    f_{abc}^\text{exp} = \exp(x(a + x(b + x c)),
\quad\text{where}\quad
    x = \rr{\sin^2\theta}.
}

\begin{figure}
    \centering
    \includegraphics[width=\linewidth]{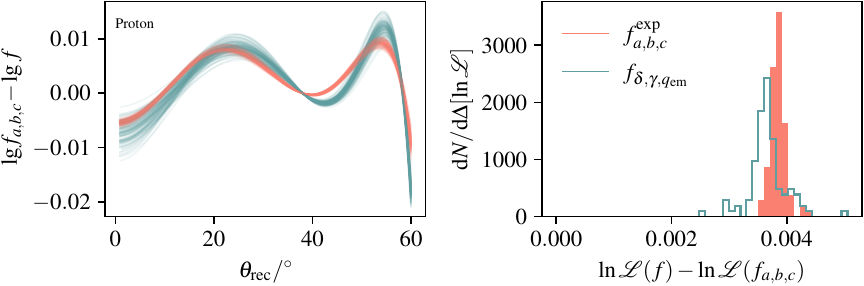}
    \caption{\emph{Left:} Difference in the fitted attenuation functions $f_{abc}^\text{exp}$ and $f_{\delta,\gamma,q_\text{em}}$ compared to the cubic function $f_{abc}$ for each of the 100 samples as a function of the zenith angle. 
    \emph{Right:} Difference in the log-likelihood for the attenuation functions compared to the cubic function.
    Higher values suggest the respective function to result in a higher likelihood and therefore suggest a better description of the data.
    A distribution around zero would suggest similar performance.}
    \label{f:compare-likelihood}
\end{figure}

To compare the different forms for the attenuation function, we draw 100 samples with 5000 events above the respective energy threshold and perform the fits.
In \cref{f:compare-likelihood} \emph{(left)} the difference between the respective attenuation form and the simple cubic function is shown.
For each of the samples the difference in the log-likelihoods is also computed and shown in \cref{f:compare-likelihood} \emph{(right)}.
We observe that the simplified two component model and the exponential cubic function describe the data consistently better than the cubic function.
It seems that the exponential cubic function however also outperforms our two component model.
This can be caused by to many simplifications in our model, or correlations in the fit parameters.
Please note that these conclusions may depend on the chosen hadronic interaction model as well.

\section{Comparison of different primary particles}

\begin{figure}
    \centering
    \includegraphics[width=\linewidth]{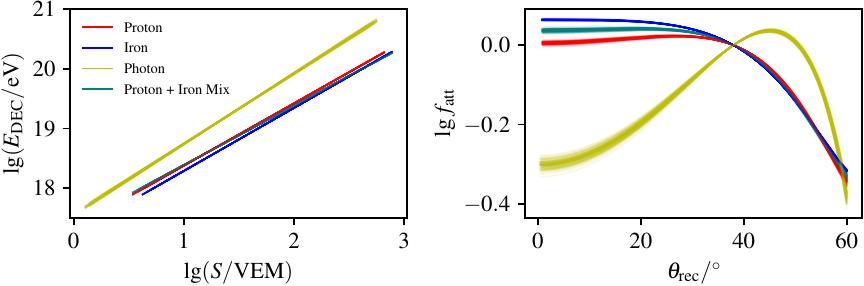}
    \caption{Energy calibration (left) and attenuation function (right) for 100 fits of the simplified two component model to showers induced by the three different primary particles and the mix of proton and iron.}
    \label{f:compare_primaries}
\end{figure}

In \cref{f:compare_primaries} we compare the different attenuation behavior of showers induced by protons, iron or photons for the simplified two component model.
As expected we observe that for lighter particles same energies are mapped to lower signals.
Moreover, as photons and high energy vertical protons will develop their maximum below the detector plane, the attenuation functions increase with zenith angles at low angles before they reach the region of exponential attenuation.
The teal curves correspond to a mixed composition linearly going from pure proton at $10^{18.5}$\,eV to pure iron at $10^{20.2}$\,eV.
As expected, we observe the fit for this mix to lay between the curves for pure proton and iron.

\begin{figure}
    \centering
    \includegraphics[width=\linewidth]{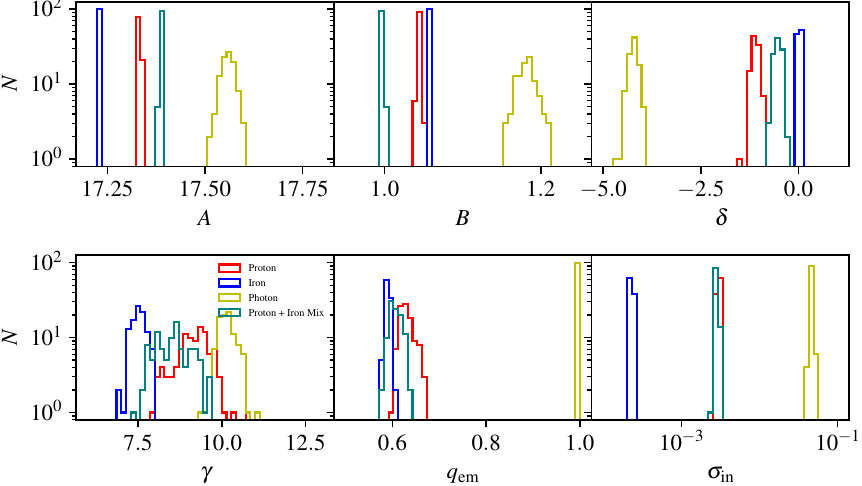}
    \caption{Distribution of the fitted parameters of the simplified two component model for 100 fits to showers induced by different primary particles.}
    \label{f:compare_parameters}
\end{figure}

An advantage of the simplified two component model is the relation of the fit parameters to physical quantities, see \cref{f:compare_parameters}.
Looking at the distributions of $q_\text{em}$, we observe the expected ordering from heavy to light and an almost pure electromagnetic composition for the photon induced shower.
As expected the intrinsic shower-to-shower fluctuation are orders of magnitudes higher for photon induced showers and smallest for iron induced showers due to superposition in the shower development. 

Looking at the distribution of the parameter $B$ we see a small deviation from $B=1$ for proton and iron and a distribution peaking around a larger value for photons.
As we expect the deviation of $B$ from 1 to be caused by the elongation rate, i.e.\ $X_\text{max}$ increasing with energy, this corresponds to photon-induced showers having a larger elongation rate than showers induced by hadrons.
For the mixed composition evolving from pure proton to pure iron the elongation rate is effectively reduced, as iron-induced showers have on average lower $X_\text{max}$ values.
Therefore also the fitted value for $B$ moves closer to 1.

\section{Conclusions}

We introduced a physically motivated model for shower attenuation, describing the muonic and electromagnetic components with a Gaisser-Hillas function.
While this form captures key features of the data, its parameters provide insight into shower dynamics but often act as effective, rather than strictly physical, quantities.
Introducing $\sigma_\text{in}$, a fit parameter representing the intrinsic shower-to-shower fluctuations, significantly reduces fit bias.
Nonphysical results when extrapolating outside of the specific fit range can be avoided by using either the derived form or an exponential ansatz.

\paragraph{Acknowledgments}

The authors are very grateful to the Pierre Auger Collaboration for providing the tools necessary to perform the simulations for this contribution and for all the fruitful discussions.

\let\oldbibliography\thebibliography
\renewcommand{\thebibliography}[1]{%
  \oldbibliography{#1}%
  \setlength{\itemsep}{0pt}%
}
{\small
\bibliographystyle{JHEP_mod}
\bibliography{bibfile}}

\providecommand{\href}[2]{#2}\begingroup\raggedright\begin{thebibliography}{1}

\bibitem{newton}
D.~Newton, J.~Knapp and A.~Watson,
  \href{https://doi.org/10.1016/j.astropartphys.2006.08.003}{\emph{Astropart.\
  Phys.} {\bfseries 26} (2007) 414}.

\bibitem{corsika}
D.~Heck, J.~Knapp, J.~Capdevielle et~al., {\emph{Report FZKA} {\bfseries 6019}
  (1998) }.

\bibitem{epos}
T.~Pierog, I.~Karpenko, J.M.~Katzy et~al.,
  \href{https://doi.org/10.1103/PhysRevC.92.034906}{\emph{Phys. Rev. C}
  {\bfseries 92} (2015) 034906}
  [\href{https://arxiv.org/abs/1306.0121}{{\ttfamily 1306.0121}}].

\bibitem{offline}
S.~Argiro, S.L.C.~Barroso, J.~Gonzalez et~al.,
  \href{https://doi.org/10.1016/j.nima.2007.07.010}{\emph{Nucl. Instrum. Meth.
  A} {\bfseries 580} (2007) 1485}
  [\href{https://arxiv.org/abs/0707.1652}{{\ttfamily 0707.1652}}].

\bibitem{vasylyev}
D.~Vasylyev, {\emph{Earth Planets Space} {\bfseries 73} (2021) 112}.

\bibitem{gh}
T.K.~{Gaisser} and A.M.~{Hillas}, {\emph{Proc.\ Int.\ Cosmic Ray Conf.}
  {\bfseries 8} (1977) 353}.

\bibitem{atmo}
{\scshape Pierre Auger} Collaboration,
  \href{https://doi.org/10.1016/j.astropartphys.2009.06.004}{\emph{Astropart.\
  Phys.} {\bfseries 32} (2009) 89}.

\bibitem{likelihood}
H.~Dembinski, B.~K\'egl, I.~Mari\c{s} et~al.,
  \href{https://doi.org/10.1016/j.astropartphys.2015.08.001}{\emph{Astropart.\
  Phys.} {\bfseries 73} (2016) 44}.

\end{thebibliography}\endgroup

\clearpage

\end{document}